# In situ monitoring of atomic layer epitaxy via optical ellipsometry

F. Lyzwa[1,2], P. Marsik[2], V. Roddatis[3], C. Bernhard[2], M. Jungbauer[1] and V. Moshnyaga[1]

[1]I. Physikalisches Institut, Georg-August-Universität Göttingen, Friedrich-Hund-Platz 1, 37077 Göttingen, Germany
[2]University of Fribourg, Department of Physics and Fribourg Center for Nanomaterials, Chemin du Musée 3, 1700 Fribourg, Switzerland
[3]Institut für Materialphysik, Georg-August-Universität-Göttingen, Friedrich-Hund-Platz 1, 37077 Göttingen, Germany

We report on the use of time-resolved optical ellipsometry to monitor the deposition of single atomic layers with subatomic sensitivity. Ruddlesden-Popper thin films of $SrO(SrTiO_3)_{n=4}$ were grown by means of metalorganic aerosol deposition in the atomic layer epitaxy mode on $SrTiO_3$(100), LSAT(100) and $DyScO_3$(110) substrates. The measured time dependences of ellipsometric angles, $\Delta(t)$ and $\Psi(t)$, were described by using a simple optical model, considering the sequence of atomic layers SrO and $TiO_2$ with corresponding bulk refractive indices. As a result, valuable online information on the growth process, the film structure and defects were obtained. Ex situ characterization techniques, i.e. transmission electron microscopy (TEM), X-ray diffraction (XRD) and X- ray reflectometry (XRR) verify the crystal structure and confirm the predictions of optical ellipsometry.

1. **Introduction**

Nanoscience and modern material physics are based on the growth and study of thin films with thicknesses spreading down to one atomic layer (monolayer, ML). Several phenomena of great current interest are coupled to this scale, e.g. high mobility electron gas at the interface between the insulator $LaAlO_3$ and $SrTiO_3$ (STO) [1], high temperature superconductivity in a FeSe monolayer on a STO substrate [2,3], interfacial ferromagnetism [4] and strain-engineering to obtain ferroelectric properties in STO [5]. To fabricate those systems several deposition techniques can be employed, including molecular-beam epitaxy (MBE), pulsed laser deposition (PLD) and chemical vapor deposition (CVD). Therefore, the growth monitoring process itself is of great importance in order to monitor and to confirm the successful film growth and, finally, to gain information on the electronic properties, defect formation, interfacial reconstructions etc. already during the growth itself.

Different methods have been used to monitor the growth of thin films. The most common approach is "Reflection High-Energy Electron Diffraction" (RHEED), in which the interference pattern of reflected electrons upon the film surface is detected. Complete unit cells (u.c.) or sometimes half u.c. can be observed, depending on the material system. However, one can only distinguish between the different growth modes such as island or layer-by-layer growth, but hardly identify what kind of material is deposited. This method is also limited to a low partial pressure of the gas atmosphere or the number of aerosol particles inside the deposition chamber. Optical probe techniques such as reflectance-difference spectroscopy (RDS), p- polarized reflectance spectroscopy and spectral ellipsometry (SE) [6] are advantageous because of their noninvasive and nondestructive character. With these spectroscopic techniques, information about the film thickness, optical properties and surface morphology can be extracted.

Nevertheless, the reflected signal is usually not detected in a continuous mode during the film growth, but rather after each growth step. Real-time monitoring for the growth of semiconductors was previously reported e. g. by Zettler J-T *et al* [7], by means of SE in the rotating analyzer ellipsometer (RAE) configuration with a monochromator placed in the beam path, suffering from a relatively low signal-noise ratio (SNR). Using a laser as the probe light improves the SNR, as was shown e.g. by Zhu X D *et al* where they monitored the growth of a STO thin film on a STO substrate via "Oblique-Incidence Reflectance-Difference" (OIRD) technique [8]. Comparing the optical signal with RHEED, a good agreement between these two monitoring techniques for an interrupted growth cycle and full u.c. of STO was established. In later publications they developed a relatively complex mean-field theory to describe the optical response [9, 10]. In our previous study [11] we have shown, that by using an ellipsometry setup in the <u>P</u>olarizer-Photoelastic <u>M</u>odulator-<u>S</u>ample-<u>A</u>nalyzer (PMSA) configuration one can detect the evolution of the optical ellipsometry signal during the deposition of atomic layers in real-time. Following those statements, the question arises 1) whether we can monitor/control the growth of oxides in general down to the submonolayer level in real-time, 2) what kind of model is necessary to describe the growth process and 3) which further information can be obtained from modeling the data?

Here, by using the above mentioned experimental setup, we performed a detailed study of the heteroepitaxial growth of the Ruddlesden-Popper (RP) series Sr-O[Sr-O/Ti-O$_2$]$_{n=4}$ in the atomic layer epitaxy (ALE) mode. Moreover, the change of the ellipsometric signal during the ALE growth monitoring was described by a relatively simple model considering optical constants of neutral sub-monolayers of Sr-O and Ti-O$_2$, as building blocks of a RP system (see Fig. 2c). The information obtained from in situ ellipsometry was verified by ex situ characterization of the crystal structure by means of transmission electron microscopy (TEM), X-ray diffraction (XRD) and reflection (XRR).

## 2. Sample preparation and ellipsometry model

All samples were fabricated via metalorganic aerosol deposition (MAD) technique [12]. In order to grow atomic layers of Sr-O and Ti-O$_2$ on top of each other, the precursors Sr(acetylacetonate)$_2$ (P$_{Sr}$) and Ti(isoprop)$_2$(tetramethylheptanedionate)$_2$ (P$_{Ti}$) were first dissolved separately in dimethylformamide (DMFA). Each precursor solution was sequentially sprayed by a pneumatic nozzle (droplet size ~20 μm) using compressed air onto a heated substrate, T$_{sub}$~900°C, where a heterogeneous pyrolysis reaction on the substrate surface occurs. The volumes of each precursor solution and, hence, the thickness of layers were controlled in separate pulses via a precision pump system, resulting in deposition rates of r=0.06-9.5 ML/s. In between the Sr-O and Ti-O$_2$ pulses (atomic layers) we chose a delay time of 6-8 seconds. Prior to deposition, the STO substrates were Ti-terminated in order to remove the top Sr-O layer at the substrate surface, as described in [13].

The in situ growth control was performed by means of an optical ellipsometry setup in the PMSA-configuration as described in detail in Ref. [11]. A HeNe-Laser beam (λ=632.8nm) was aligned close to the Brewster angle (φ$_B$=62°) with respect to the substrate normal. Polarizer and analyzer angles were both set to 45°. The polarization of the light was modulated at a frequency of ω=50 kHz by means of a photoelastic modulator placed between the polarizer and the sample. The resulting intensity modulation was detected with a photodiode via a lock-in technique and was used to derive the ellipsometric angles $\Psi$ and $\Delta$, given by the ratio $\rho$ of the Fresnel coefficients $r_{(s,p)}$:

$$\frac{r_p}{r_s} = \rho = \tan\Psi \cdot e^{i\Delta} \quad \Rightarrow \quad \Delta = \tan^{-1}\left(\frac{Im\,\rho}{Re\,\rho}\right) \text{ and } \Psi = \tan^{-1}(|\rho|) \qquad (1)$$

The time-resolved optical response ($\Delta, \Psi$) of a thin film during the growth as a function of time or film thickness was theoretically described by using a standard transfer matrix formalism (for s- and p- components), which couples the incidence ($i$), reflected ($r$) and transmitted ($t$) amplitudes of the electric field vector $E$ for an isotropic, homogeneous sample [14]:

$$\begin{bmatrix} E_i \\ E_r \end{bmatrix} = \widehat{M} \cdot \begin{bmatrix} E_t \\ 0 \end{bmatrix} \qquad (2)$$

Imagine now, we deposit a layer A with a complex refractive index, $\tilde{n}_A$, and thickness, $d_A$, on top of a substrate (see Fig. 1). We can express $E$ at the interface ambient air (amb)/layer A by a 2x2 matrix $R_{amb/A}(r_{(s,p)}, t_{(s,p)})$, composed by the Fresnel coefficients. The wave travelling through the layer A is expressed also by a 2x2 matrix, $\phi_A(\tilde{n}_A, \Theta_2, \omega, d_A)$, with the angle $\Theta_2$ under which the wave is traveling through the layer and its energy $\omega$. The total optical response of the material is described by M$_1$, which is determined by the contribution of the substrate and the layer A. Using $R_{A/amb} \cdot R_{amb/B} = R_{A/B}$ we can describe a free standing layer A and derive:

$$\begin{bmatrix} E_i \\ E_r \end{bmatrix} = \underbrace{R_{amb/A} \cdot \phi_A \cdot R_{A/sub}}_{M_1} \cdot \begin{bmatrix} E_t \\ 0 \end{bmatrix} = \underbrace{R_{amb/A} \cdot \phi_A \cdot R_{A/amb}}_{M_A} \cdot \underbrace{R_{amb/sub}}_{M_0} \cdot \begin{bmatrix} E_t \\ 0 \end{bmatrix} \qquad (3)$$

We assume $d_{sub} \to \infty$, since we expect a negligible contribution from the backscattered light also due to the fact that the substrates are only one-side polished.

An additional layer B with properties $(ñ_B, d_B)$ creates a different set of matrices $(R_{amb/B}, ϕ_A, R_{B/amb})$, and one obtains the response $M_2 = M_B \cdot M_1$.

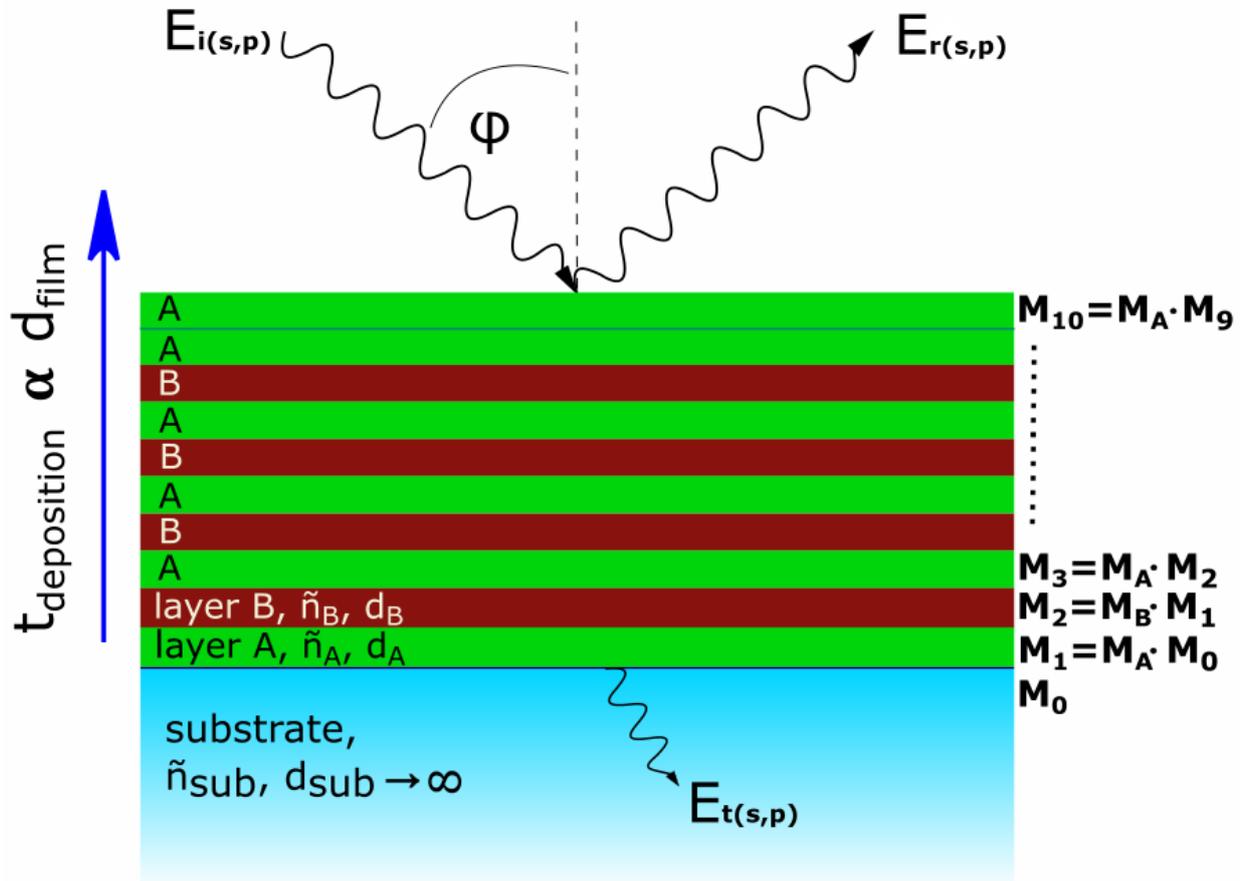

**Figure 1.** Schematic of the sample and the ellipsometry modeling. The film shown consists of 4u.c. of STO with an SrO-extra layer on top.

According to the RP series, we alternate $M_A$ and $M_B$ and calculate the optical response as

$$\hat{M} = M_1 \cdot M_2 \cdot M_3 \cdot … \cdot M_j \quad (4)$$

for *j* numbers of deposited layers. Using Eq. (1) we obtain the theoretical values $(Δ, Ψ)$ for any film thickness/deposition time, since the Fresnel coefficients $r_{(s,p)} = \frac{E_{r(s,p)}}{E_{i(s,p)}} = \frac{m_{21(s,p)}}{m_{11(s,p)}}$ are described by the elements of the matrix $\hat{M}$.

In order to confirm the presence of the Ruddlesden-Popper structure in the grown films, an additional ex situ structural characterization was done by means of XRD in the θ-2θ geometry and XRR, using a D8 Advance diffractometer from Bruker AXS. The analysis of the atomic structure of the films was performed by High-Resolution Scanning Transmission

Electron Microscopy (HR-STEM) and Electron Energy Loss Spectroscopy (EELS). The TEM samples were prepared by mechanical polishing followed by Ar⁺ ion milling at 4 kV down to 0.5 kV until electron transparency was achieved. HR-STEM images were recorded with parallel beam illumination using a FEI Titan 800-300 environmental TEM operated at 300 kV acceleration voltage. The microscope is equipped with a Gatan Image Filter Quantum 965ER. Spectrum Images (SI) were collected to confirm chemical composition of the RP planar defects.

## 3. Results and discussion

In Fig. 2a the time evolution of the measured ellipsometric angle, $\Delta$, for a RP-STO film on a STO-substrate is presented. The inset shows an enlarged view of the time evolution during the deposition of a single Ti-O$_2$/Sr-O cycle. Note, that after the deposition of one Ti-O$_2$ atomic layer a 6 sec pause in the growth process (marked in orange color) was applied, before a Sr-O atomic layer is deposited. During this time, even though no further material is deposited, a change in $\Delta$ can still be seen, possibly due to the growth kinetics and/or surface reactions.

**Table 1.** Model parameter used to describe the development of the ellipsometry signal. The refractive indices for bulk SrO and TiO$_2$ were taken at a wavelength of λ=589 nm from the literature [15]. The values for atomic layer thicknesses were estimated to be $d_{SrO} = 0.21$nm and $d_{TiO_2} = 0.19$nm. The deviation of the adjusted values compared to the bulk values are shown in brackets.

| Substrate | $n_{SrO,\text{bulk}}$ /$n_{TiO_2,\text{bulk}}$ | $n_{SrO,\text{adj.}}$/$n_{TiO_2,\text{adj.}}$ for d < 50nm | $n_{SrO,\text{adj.}}$/$n_{TiO_2,\text{adj.}}$ for d < 12.5nm | $n_{SrO,\text{adj.}}$/$n_{TiO_2,\text{adj.}}$ for d < 7.1nm |
|---|---|---|---|---|
| **STO** $n = 2.34, k = 0$ | 1.871 / 2.612 | 1.869 / 2.645 (-0.1% / 1.2%) | - | - |
| **LSAT** $n = 2.054, k = 0$ | -«- | 1.83/2.61 (-2.2% / -0.1%) | 1.96/2.68 (4.8% / 2.6%) | - |
| **DyScO$_3$** $n = 2.1, k = 0$ | -«- | 1.85/2.62 (-1.1% / 0.3%) | 1.91/2.67 (2.1% / 2.2%) | 2.02/2.74 (8.0% / 5.0%) |

The corresponding simulation as a function of the film thickness is shown in the bottom panel of Fig. 2a. As start parameters for the modeling we used room temperature values of the refractive index, ñ, of the bulk SrO and TiO$_2$ (rutile) materials [15]. These values were then slightly adjusted to better reproduce the experimental data, with final deviations of -0.1% for SrO and 1.2% for TiO$_2$. The small differences can be partly understood in terms of a temperature effect, since the film was grown at elevated temperatures. The value of ñ for the substrate has been derived from the measured $\Delta, \Psi$-values before the deposition started. All values, as well as the atomic layer thicknesses, are listed in Tab. 1. Note that the reduction of $\Delta$ at a higher film thickness arises from the interference of the laser beam light which is reflected from the film/amb and the substrate/film surfaces. This interference effect is reproduced by our simulations for the ideally grown film.

Two major deviations of the experimental data from the ideal model growth can be observed: 1) During the growth of a 4 u.c. stack of STO the phase shift angle $\Delta$ changes by about $\Delta^{4STO} = 0.36°(2)$, which is rather reproducible throughout the whole measurement. The only exception occurs for the very first stack at the start of the deposition (red square), for which the amplitude of $\Delta^{4STO} = 0.18(2)$ is reduced by a factor of two, suggesting that only half of Sr-O and Ti-O$_2$ atomic layers are deposited here. Correspondingly, in the simulation for the first stack only half of the atomic layer thicknesses were considered, i.e. $d_{SrO} \rightarrow \frac{d_{SrO}}{2}$, $d_{TiO_2} \rightarrow \frac{d_{TiO_2}}{2}$. We assume that this slower growth at the very beginning of the deposition might be related to a lower surface mobility and reduced adsorption of the reactants on the bare STO substrate. This leads to a incomplete first layer with Sr-O islands separated by gaps, closing during the next Ti-O$_2$ pulse. An additional Sr-O layer compensates this process and increases the adsorption; 2) For the first 10-12 nm of the film – this region will be called transition zone (TZ) from now on – the experimental data deviates slightly from the modeled curve in the overall shape as well as in a small horizontal drift during a 4 u.c.-STO repetition. A reason for this difference in ñ can be small deviations from the ideal monolayer doses, since the stoichiometry is not necessarily constant during the entire growth process. One possible explanation is that SrO diffuses into the substrate and therefore changes the ideal stoichiometry for the first layers. Another reason can be the formation of defects, which disturb the growth of RP-STO film with a well-defined structure. After overcoming the TZ (d>12nm) the experimental data are almost in a perfect agreement with the simulation. This will be further discussed below by presenting the strained RP- films and the corresponding TEM images.

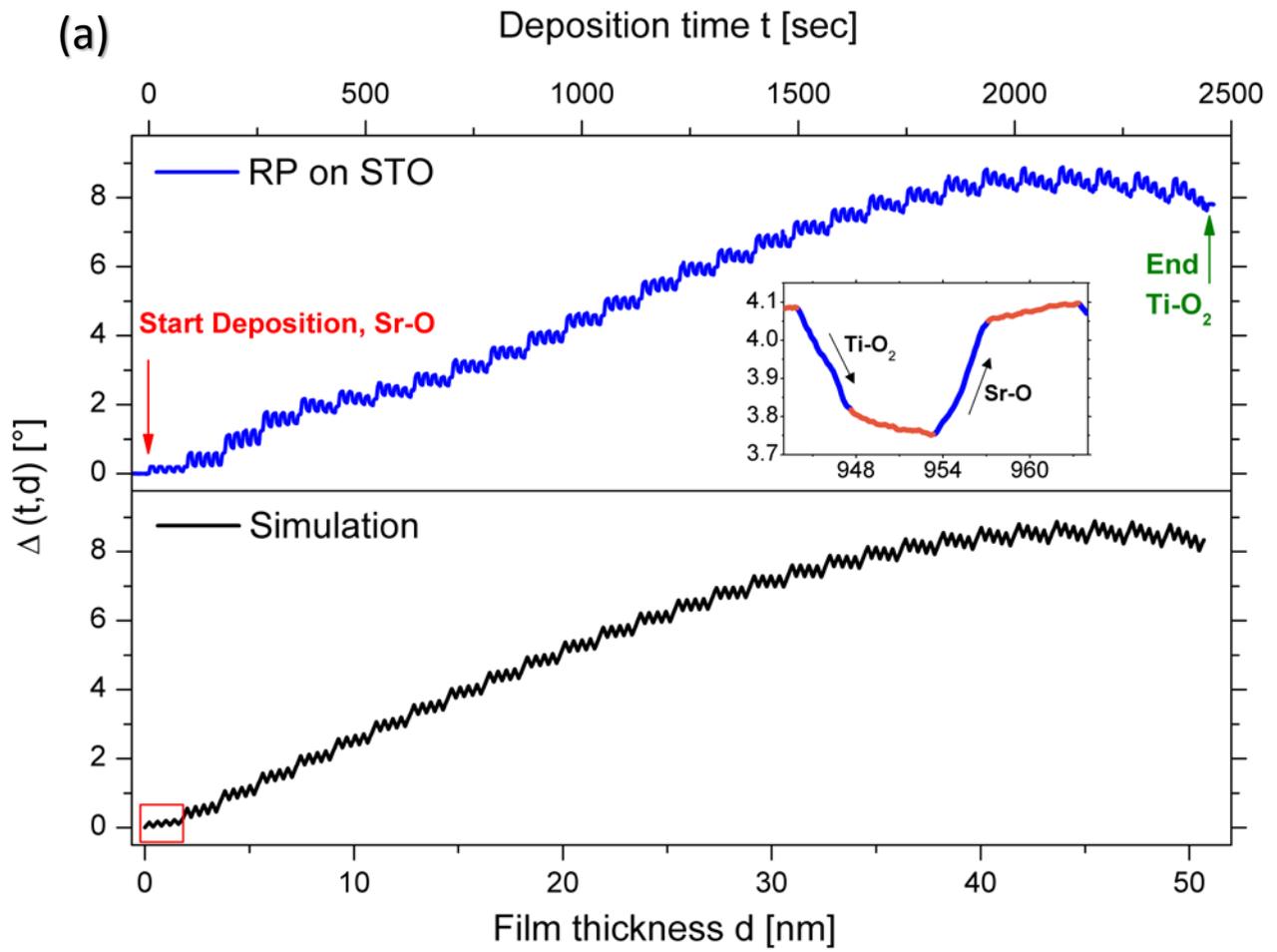

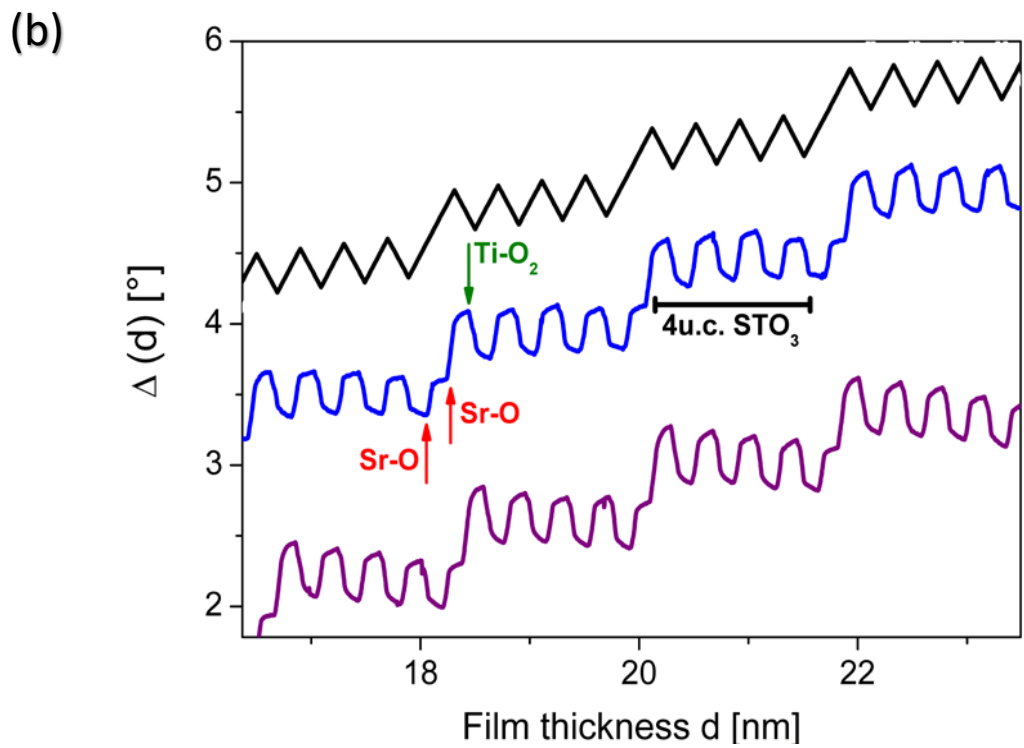

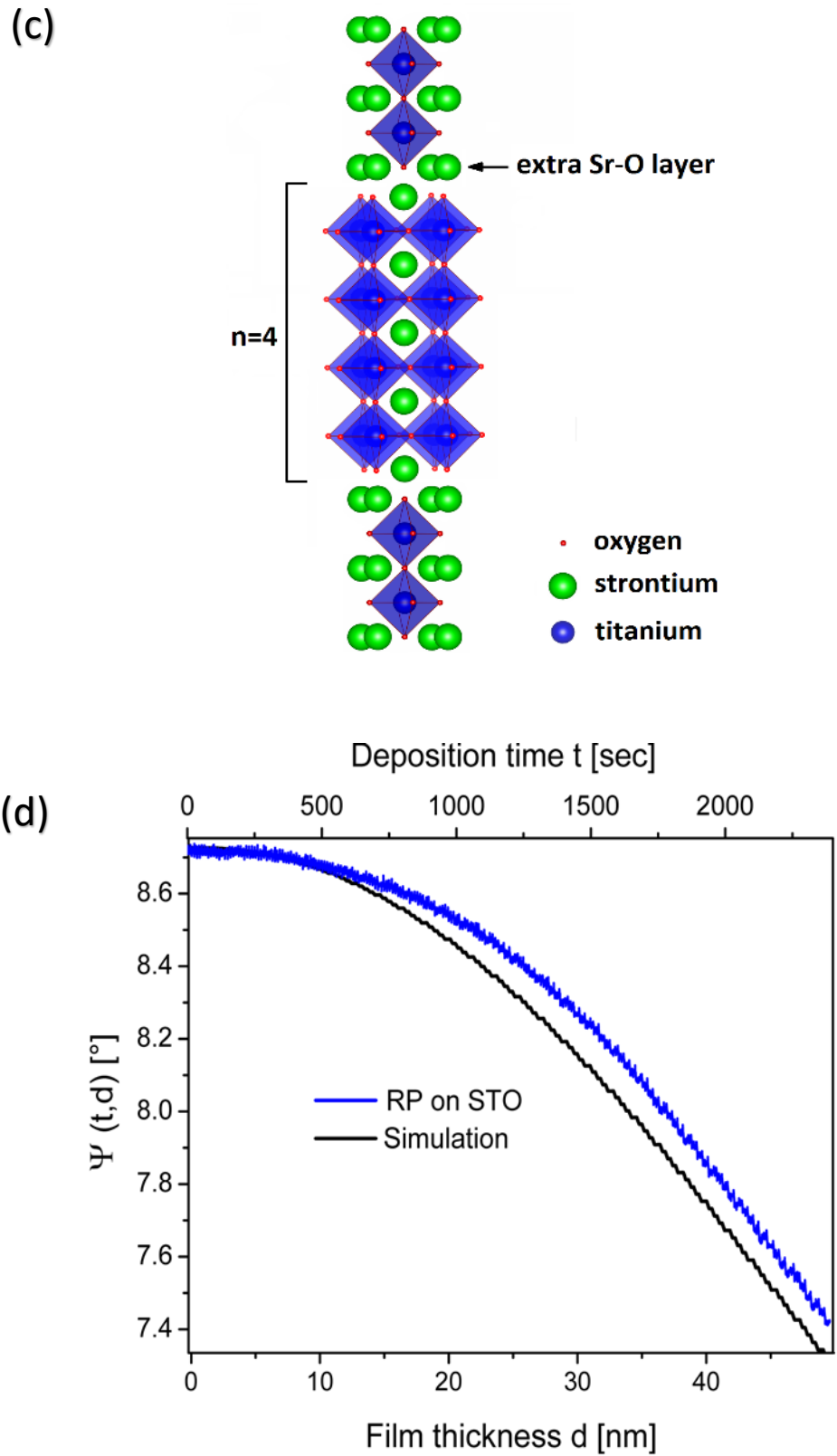

**Figure 2**. (a) Top-panel: Time evolution of the ellipsometric angle, $\Delta$, as measured during the growth of a RP-film on STO (001) with N=28 repetitions (blue curve). The inset details the evolution during the deposition of one Sr-O and one Ti-O$_2$ atomic layer. The pause in between each of the pules is shown in orange color. Bottom-panel: Calculated evolution of $\Delta$ as a function of the film thickness (black-line). (b) Magnified view of (a), including an additional RP-STO film on a STO substrate for which the stochiometry is not correct (purple). (c) Sketch of the Ruddlesden-Popper Structure with $n = 4$ u.c. of SrTiO$_3$ separated by single SrO atomic layers. (d) Corresponding experimental and simulated time evolution of $\Psi$.

Figure 2b shows a magnified view for a part of the growth process in (a). In addition, the figure shows the data for a second film (purple color) which was grown with the same deposition parameters as for the one shown by the blue line, with the only exception of a 13% lower Sr-content in the solvent; this corresponds to the ideal stoichiometry value to grow $SrTiO_3$ films by conventional MAD. This deviation leads to clearly visible changes of the ellipsometry data, i.e. it results in a horizontal drift of $\Delta$ between the consecutive growth of a double Sr-O layer. The ex situ XRD confirms the absence of the desired RP-phase for this "poorly" grown film (not shown here). Those non-stoichiometric growth conditions within ALE are also reported for Ruddlesden-Popper structures grown by MBE [16].

The corresponding evolution of the ellipsometric angle, $\Psi$, for the high quality RP-STO film on a STO substrate is displayed in Fig. 2d. The atomic-layer oscillations are clearly resolved in the $\Psi(t)$ curve, although their amplitude is significantly smaller than in the $\Delta(t)$ curve as is expected since the measurements are performed close to $\varphi_B$ of STO. Simulation and experiment follow the same trend, even though there is less agreement than for $\Delta$ (see Fig. 2a). A large deviation was especially found for the time (thickness) of t>500 s (d>10 nm), while for a thinner film the deviation is rather small.

The time evolution of $\Delta$ for the strained RP-films that are grown on the LSAT and the $DyScO_3$ substrates are shown in Figure 3 a, b. The difference between the refractive indices of these substrates and the RP-film are larger than for the STO substrate, thus, leading to more pronounced interference effects (minima in $\Delta$(t,d)) already for thinner films, d~15-20 nm. Nevertheless, for the strained films it is more difficult to obtain a good agreement between the model and the experimental data. The first ~12 nm of the film seem to have a much higher refractive index than the rest of the RP-film. This effect of a transition zone (TZ) we already observed in an RP film grown on a lattice matched STO substrate (see Fig. 2). In contrast, for strained RP-films the whole deposition was not possible to be simulated by using only by one ñ-value for each Sr-O and Ti-$O_2$ atomic layer. Therefore, we described the TZ with one or two additional refractive index(cis) for the case of a LSAT and $DyScO_3$ substrate, respectively (see Tab. 1). For film thicknesses $d$ > TZ the refractive indices used for the simulation match the bulk values taken from the literature better than for the case d < TZ, suggesting a lower film quality in the first part (d<TZ) of the film. Note, that within the TZ the refractive index changes probably not in a stepwise manner but rather gradually. Nevertheless, our model with the ñ- values shown in Tab. 1 describes the experimental data quite well. The differences in the used ñ for different substrates can be understood in terms of biaxial strain effect, i.e. tensile for DyScO$_3$ and compressive for LSAT. The strain is known to affect the band structure and, therefore, the refractive index of a given material, as was shown for strained STO thin films [17]. In the corresponding $\Psi(t)$ curves of RP-films on LSAT (Fig. 3c) and DyScO$_3$ (Fig. 3d) atomic layer oscillations are observable, but are less pronounced than those in $\Delta(t)$ due to the same reason as for the RP-film on a STO substrate, described above. However, the simulated $\Psi(t)$ curves oscillate weaker than the experimental data, and it was not possible to find any simulation parameters to satisfy simultaneously the $\Delta(t)$ and $\Psi(t)$ evolution. This indicates, that even though several online information can be extracted from the measured ellipsometry data and the model accounts for most of the observed effects, nevertheless, this model seems to be not complete. The reason for that is, most probably, the fact that the Fresnel coefficients assume planar interfaces and, hence, the film roughness will not be represented by the equations used in this model. This could explain the progressively higher values of $\Psi$ in the experiment as the deposition continues and the film thickness and possibly the roughness increases.

Fig. 3e shows the $\Delta(t)$ development at the early deposition stage of different RP samples grown on DyScO$_3$ and LSAT substrates. One can see that the Δ-values of the first stack of 4 STO u.c. differ from those of the following repetitions (as was seen above in the case of a STO substrate). If we follow the Sr-O[Sr-O/Ti-O$_2$]$_{n=4}$ procedure, shown in the upper two panels of Fig. 3e, it seems that the growth of the SrO atomic layers is suppressed until the extra SrO layer is deposited. For another sample, grown on LSAT, we immediately started the deposition of doubled SrO layers (magenta colored curve) and obtained pronounced atomic layer oscillations from the very beginning of the growth. Apparently, the SrO extra layer plays an important role for the deposition of those oxides and catalyzes the growth on all used substrates. The corresponding simulation confirms this behavior.

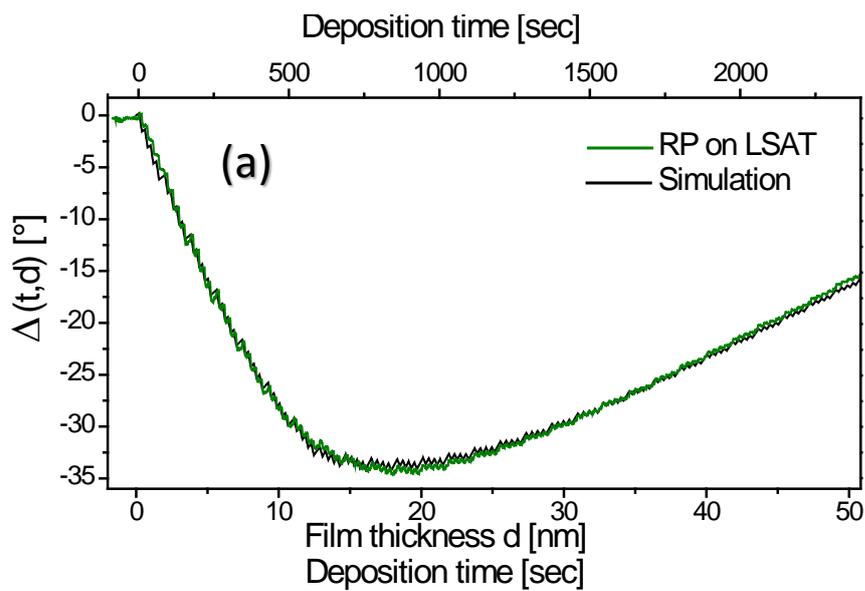
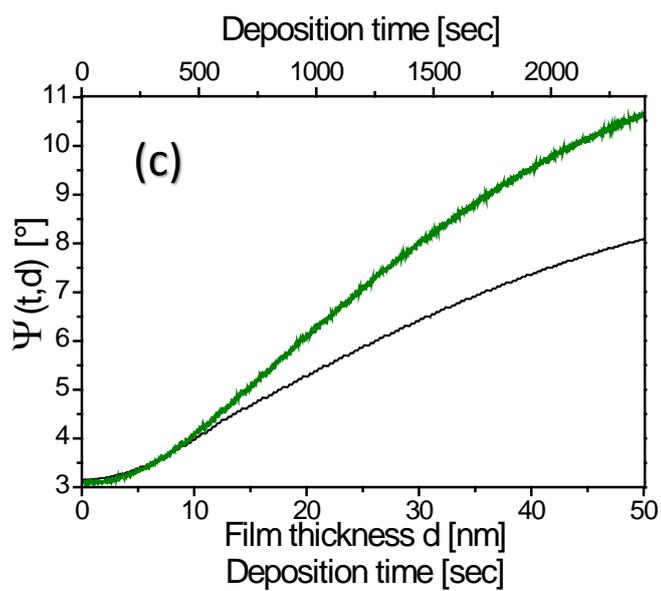
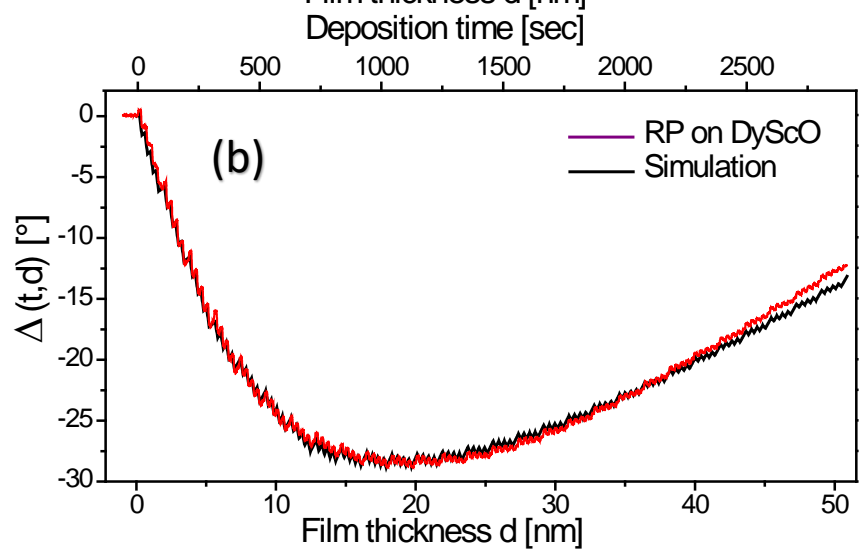
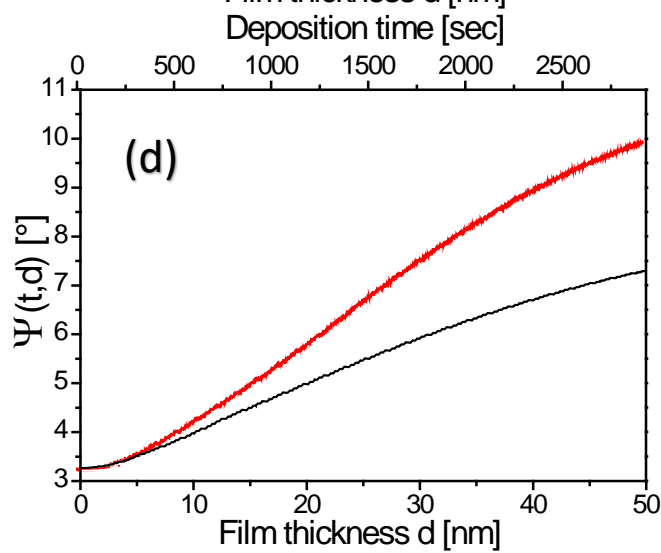

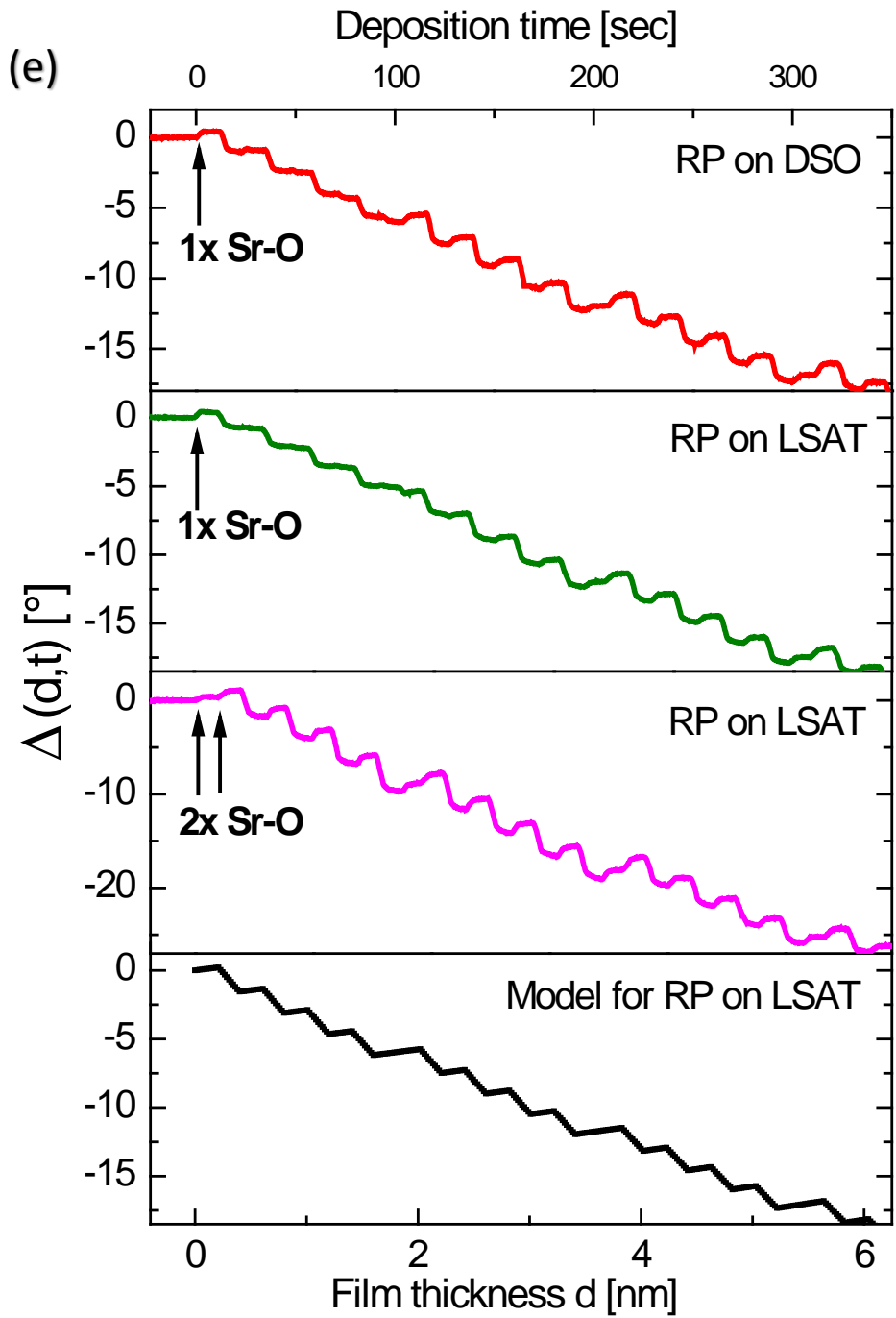

**Figure 3.** Ellipsometry signal of strained RP films on LSAT (in green, a and c) and DyScO$_3$ (red, b and d) substrates together with the calculated curves (respectively for each substrate in black color). The crucial role of an additional SrO layer at the beginning of the growth is visible in (e).

Can this ellipsometry technique be equally suitable for the in situ monitoring during the growth of strongly absorbing materials? In Fig. 4 we simulated the evolution of the ellipsometric angles $\Delta$ and $\Psi$ for a virtual Ruddlesden-Popper thin film of $Sr_2RuO_4$ with $n = 1$ (SRO) epitaxially grown on a STO substrate. For a SRO film the imaginary part of the refractive index, $k$, is significantly larger than zero at the wavelength of the laser beam, λ=632.8 nm.

For the calculation we assumed that the SRO film is grown in the ALE-mode, i.e. by means of sequential deposition of the following atomic layers: [Sr-O/Sr-O/Ru-O$_2$]$_N$. The resulting atomic layer oscillations are clearly visible in the $\Delta(d)$ curve, whereas in the $\Psi(d)$ curve they are rather faint.

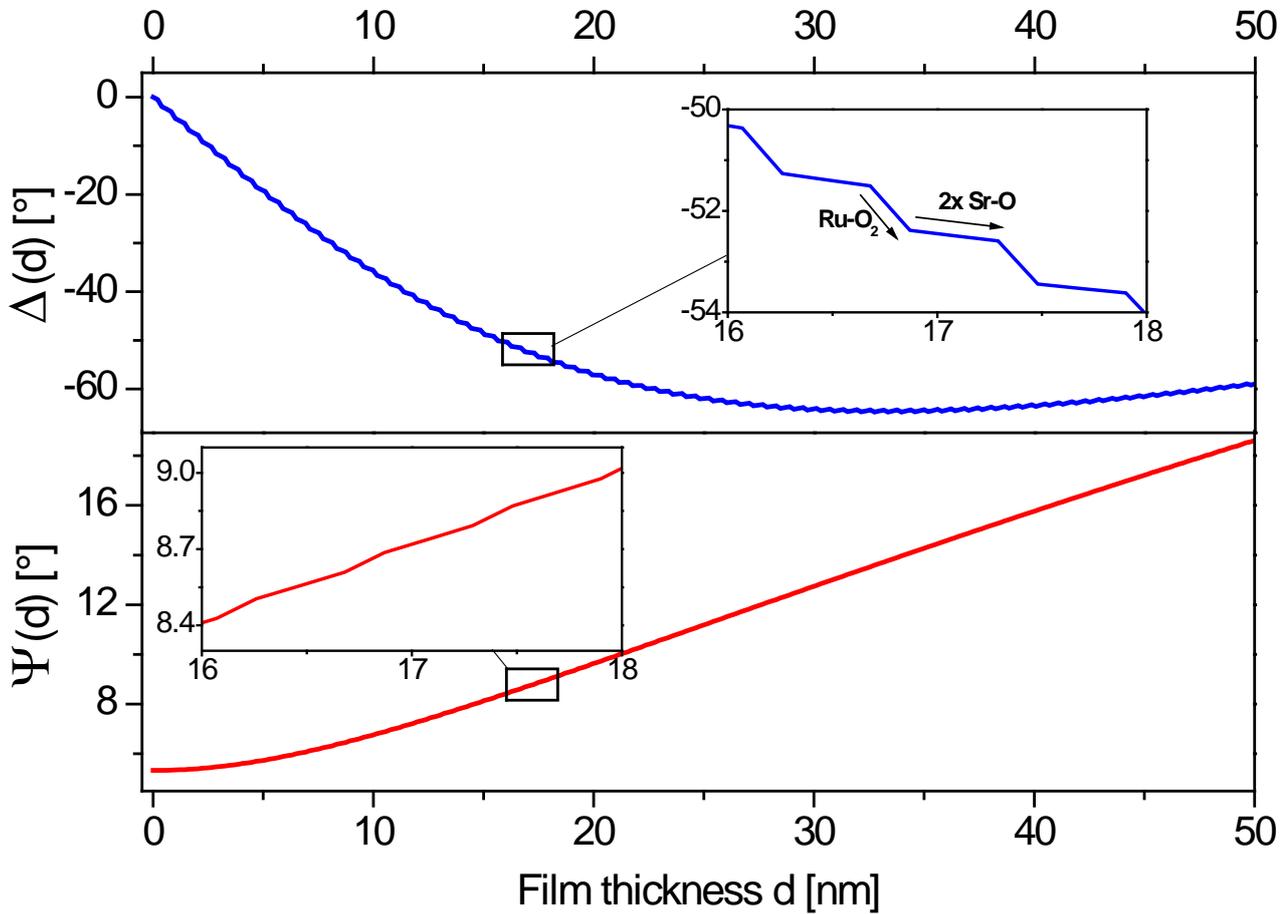

**Figure 4.** Theoretical optical response ($\Delta$, $\Psi$) of $\mathrm{Sr_2RuO_4}$ on STO(001) substrate by depositing single atomic layers of SrO and $\mathrm{RuO_2}$ (see inset), with an angle of incidence of $\varphi = 70°$. The refractive index for $\mathrm{RuO_2}$ was taken from bulk data [18] to be ñ $= 1.3 + 0.8j$.

Next, we discuss the results of the ex situ structural characterization, carried out by XRD/XRR and TEM techniques, which confirm the in situ measurements of optical ellipsometry. Figure 5a shows the XRD patterns for the samples grown on STO (blue), LSAT (green) and DyScO$_3$ (red), each with an underlying bare substrate (gray). All patterns show the desired out-of-plane epitaxy with *(00m)* x-ray peaks for the Ruddlesden-Popper phase with $n = 4$ (labeled according to the diffraction order, $m$) and confirm the crystalline growth of these films. The out-of-plane lattice parameters (c-parameters) were calculated, using Bragg's law $2d_{(hkl)} \sin \Theta = m\lambda$, from the slope of the linear fit to the different orders (see insets).

For the film on STO, the evaluated parameter, $c = 3.59(5)$nm, agrees well with previous experimental and theoretical results [11, 19]. For the film on LSAT the c-lattice constant is enlarged by about 0.6%. Such an increase is expected due to the compressive strain from the LSAT substrate with a 1% reduced in-plane lattice constant compared to STO. Similarly, for DyScO$_3$, which has a 1% larger in-plane lattice constant than STO and thus induces a tensile biaxial strain, we obtain a reduction of the c-axis lattice parameter by about -0.6%. The XRR curves for the RP-films on LSAT and DyScO$_3$ are presented in Fig. 5b and show periodic oscillations thus confirming the large scale homogeneity and the smooth surfaces of the grown RP samples. From these oscillations we derived the film thicknesses, $d = 51$nm and 54 nm, respectively for the RP-films grown on LSAT and DyScO$_3$. Note, that the contrast in the electron density between the RP STO film and the STO substrate is too low to observe these thickness oscillations.

In Fig. 5 c-e the TEM images of a RP film grown on a LSAT substrate demonstrate a well-defined epitaxial growth with chemically sharp film/substrate and internal RP interfaces. Periodical ordering of the perovskite u.c. is visible along with continuous SrO- extra layers (Fig. 5d) that are marked with white arrows. The EELS image in Fig. 5e supports the achieved $n = 4$ stacking of the Ruddlesden-Popper structure; here the Sr-atoms are colored in orange, the Ti- atoms in blue. The thickness of the film on the STO substrate could be estimated to be $d = 51(1)$nm.

Also visible are vertical SrO defects as well as areas with pure STO or RP phases with different $n$, which were previously reported for MBE and MAD grown RP systems [11, 19, 20]. These can be explained by considering the Gibb's free energy of reaction $\Delta G = \Delta H - T\Delta S$, which will be lower for a mixture of RP phases than for a pure SrO(SrTiO$_3$)$_{n=4}$ phase film, due to the higher entropy $S$ [19]. For instance, it has been shown that for $n > 3$ the RP films become thermodynamically unstable [21] and conventional solid state reaction methods fail to grow such heterostructures. Furthermore, vertically oriented SrO defects have low-energy interfaces [22].

Here, those defects mostly occur in the first ~12nm of the grown film, which matches nicely the above mentioned transition zone (TZ), observed by in situ ellipsometry (Fig. 2, 3). This region appears to be independent of the substrate origin or the film thickness. Compared to MBE grown films, for which no such TZ was detected, the growth temperature within MAD is by about $200°C$ higher, meaning that the growth kinetics might be of great importance.

We believe, that in the first few layers two-dimensional (2D) islands are formed due to a higher surface tension of the film compared to the substrate, as was previously reported by [23]. As those 2D islands grow larger and finally start to overlap, they form mixed RP-phases to decrease the Gibb's energy by the entropy term. After this TZ, a step-flow is stabilized, leading to an "atomic layer-by-atomic layer" growth mode with natural stacking of Sr-O and Ti-O$_2$ atomic layers as was designed for the RP-STO structure with $n = 4$.

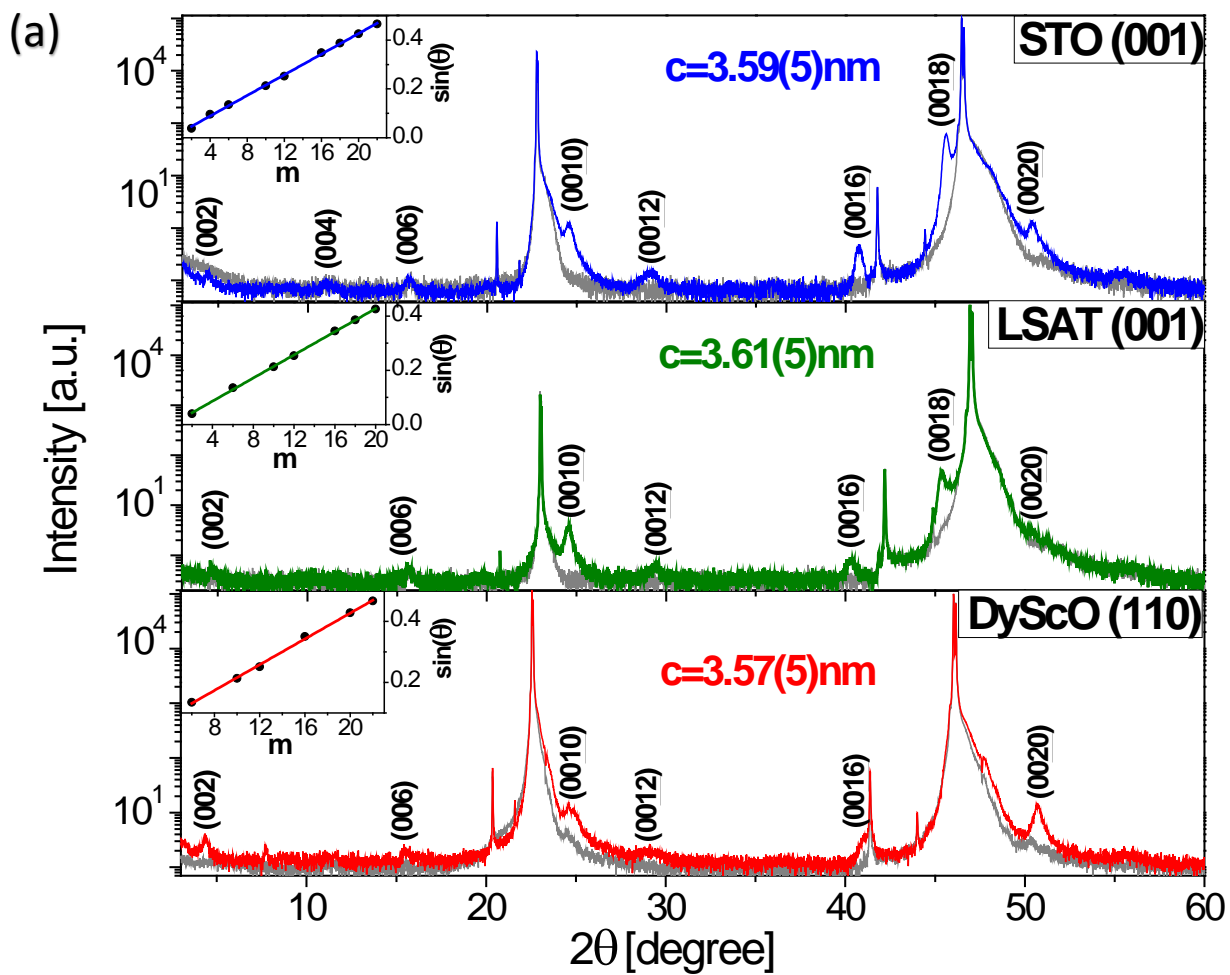

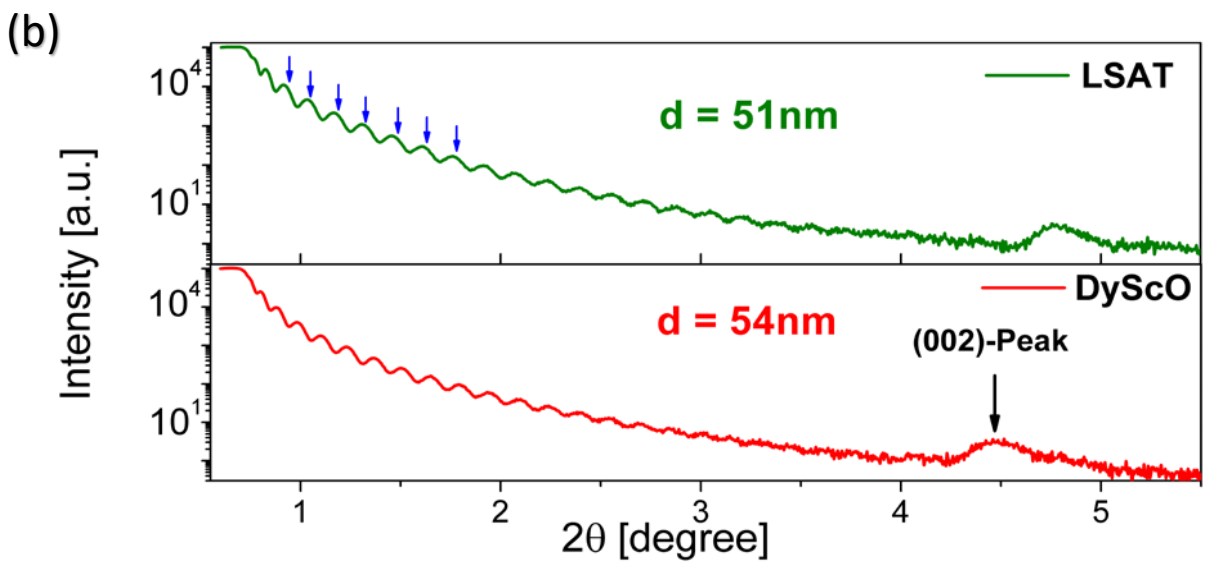

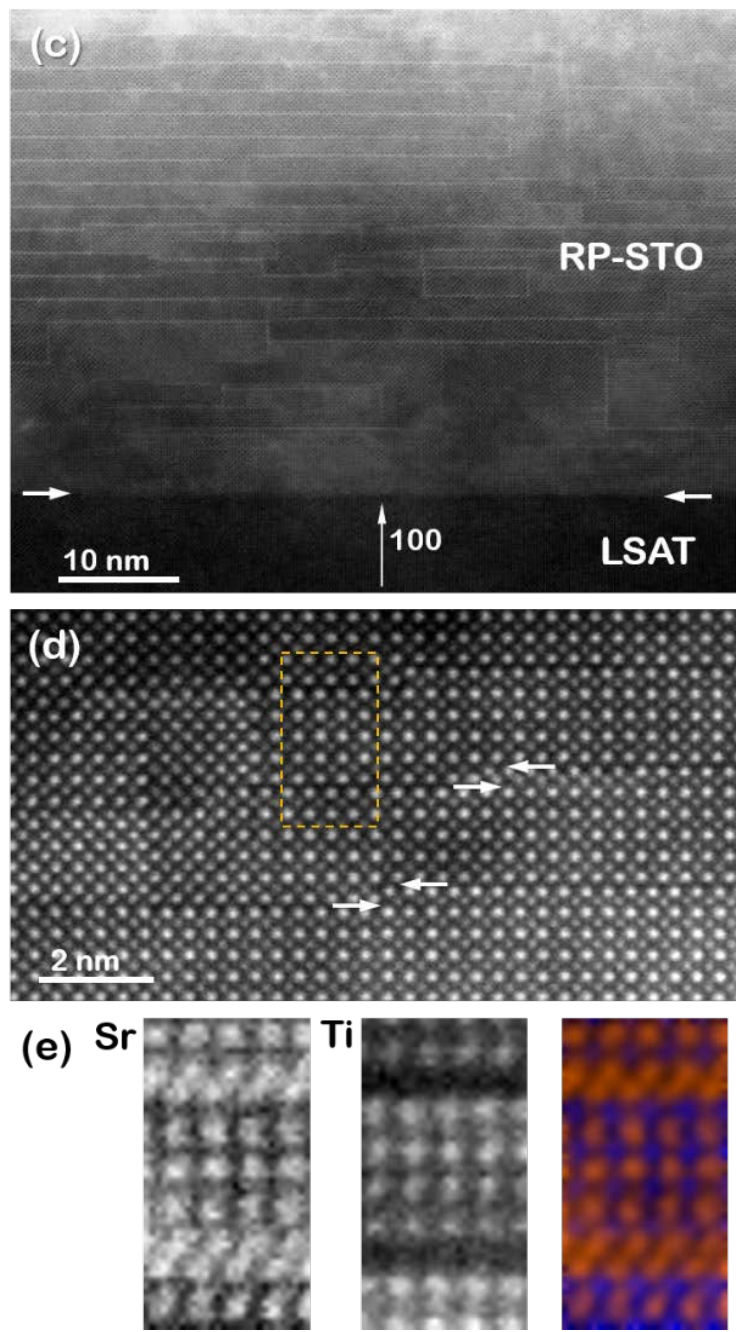

**Figure 5.** Structural analysis of RP STO films. (a) XRD pattern of RP on STO(001) (blue), on LSAT (green) and on DyScO$_3$ (red). The respective spectra of each substrate are shown in gray color. The resulting RP-peaks are clearly visible. From Bragg's law (as shown in the insets) the c-lattice parameters (out- of- plane) were calculated. (b) XRR measurements: the thicknesses of the films grown on LSAT and DyScO$_3$ were calculated from the positions of the oscillation maxima (blue arrows). (c) Overview and (d) local HR-STEM image of RP grown on LSAT. (e) EELS mapping of strontium (orange) and titanium (blue) and their superposition. Continuous SrO-extra layers are clearly visible.

## 4. Conclusions

We demonstrated the great potential of in situ ellipsometry for monitoring and studying thin film growth down to the scale of single atomic layers. Our ellipsometry setup allows monitoring the film growth down to the time scale of Δt~0.1 s, which corresponds to a film thickness resolution of Δd ~0.1 u.c.. Such an approach enables one to get an insight on the atomic layer epitaxy of complex Ruddlesden-Popper heterostructures with $n = 4$. The defect formation, i.e. vertically inter-grown SrO layers and RP phases with $n \neq 4$, were predominantly found in the first ~12 nm of the RP film and ascribed to a relatively high deposition temperature within MAD. Furthermore, we found that the SrO extra layer acts like a "growth catalyst" and induces the layered growth of these RP STO films. The optical response during the ALE growth was modeled by using refractive indices for Sr-O and Ti-$O_2$ atomic layers, which were found to deviate from the corresponding bulk data by less than 1.2 % for the unstrained film and by up to 8 % for the strained films. As the ellipsometry monitoring is not limited to film/substrate pairs (absorbing, non-absorbing and even charged atomic layers) or to the used deposition technique, we believe that our results will stimulate further research in the field of new layered materials and interface effects.

## 5. Acknowledgements

This work was financially supported by the Swiss National Science Foundation (SNF) via grant No. 200020-172611 and by the EU FP 7 Project "IFOX". We thank E Unger, S Hühn, R de Andres Prada and E Perret for helpful discussions.